\documentclass[conference]{IEEEtran}
\newcommand{\BEAS}{\begin{eqnarray*}}
\newcommand{\EEAS}{\end{eqnarray*}}
\newcommand{\BEA}{\begin{eqnarray}}
\newcommand{\EEA}{\end{eqnarray}}
\newcommand{\BEQ}{\begin{equation}}
\newcommand{\EEQ}{\end{equation}}
\newcommand{\BIT}{\begin{itemize}}
\newcommand{\EIT}{\end{itemize}}
\newcommand{\BNUM}{\begin{enumerate}}
\newcommand{\ENUM}{\end{enumerate}}

\newcommand{\BA}{\begin{array}}
\newcommand{\EA}{\end{array}}

\newcommand{\eg}{{\it e.g.}}
\newcommand{\ie}{{\it i.e.}}

\newcommand{\reals}{{\mbox{\bf R}}}

\newcommand{\diag}{\mathop{\bf diag}}

{\begin{quote}}{\end{quote}}

\makeatletter
\long\def\@makecaption#1#2{
   \vskip 9pt
   \begin{small}
   \setbox\@tempboxa\hbox{{\bf #1:} #2}
   \ifdim \wd\@tempboxa > 5.5in
        \begin{center}
        \begin{minipage}[t]{5.5in}
        \addtolength{\baselineskip}{-0.95pt}
        {\bf #1:} #2 \par
        \addtolength{\baselineskip}{0.95pt}
        \end{minipage}
        \end{center}
   \else
    \hbox to\hsize{\hfil\box\@tempboxa\hfil}
   \fi
   \end{small}\par
}
\makeatother

\newcounter{oursection}

\newcounter{lecture}

\IEEEoverridecommandlockouts
\usepackage{cite}
\usepackage{amsmath,amssymb,amsfonts}
\usepackage{algorithmic}
\usepackage{graphicx}
\usepackage{textcomp}
\usepackage{xcolor}
\usepackage{tikz}
\usepackage{array}
\usepackage{comment}
\usepackage{caption}

\usetikzlibrary{shapes, arrows}

\def\BibTeX{{\rm B\kern-.05em{\sc i\kern-.025em b}\kern-.08em
    T\kern-.1667em\lower.7ex\hbox{E}\kern-.125emX}}
\begin{document}

\title{PV Fleet Modeling via\\ Smooth Periodic Gaussian Copula}

\author{\IEEEauthorblockN{Mehmet G. Ogut$^1$, Bennet Meyers$^{2}$, and 
Stephen P. Boyd$^1$}
	\IEEEauthorblockA{$^1$ Stanford University, Stanford, CA, 94305, USA\\$^2$ 
	SLAC National Accelerator Laboratory, Menlo Park, CA, 94025, USA}}

\maketitle

\begin{abstract}
We present a method for jointly modeling power generation from a 
fleet of photovoltaic (PV) systems. We propose a 
white-box method that finds a function that invertibly maps vector time-series data to 
independent and identically distributed standard normal variables. The proposed 
method, based on a novel approach for fitting a smooth, periodic copula 
transform to data, captures many aspects of the data such as diurnal variation 
in the distribution of power output, dependencies among different PV systems, 
and dependencies across time. It consists of interpretable steps 
and is scalable to many systems. The resulting joint probability model 
of PV fleet output across systems and time can be used to generate synthetic data,
impute missing data, perform anomaly detection, 
and make forecasts.  In this paper, we explain the 
method and demonstrate these applications.
\end{abstract}

\begin{IEEEkeywords}
photovoltaic systems, photovoltaic fleet modeling, distributed power generation, power generation 
planning, forecasting, convex optimization, copula method, probability distributions, forecast 
uncertainty
\end{IEEEkeywords}
\section{Introduction}
Modeling the power output of a fleet of photovoltaic (PV) systems
is of great importance for digital operations and maintenance 
in the PV sector, which is now a multi-billion dollar industry~\cite{Fonseca2019}. 
Applications include predicting power production, 
detecting anomalies, and making informed decisions about 
when and where to send workers to service a site. 
In recent years, there has been a significant 
increase in the deployment of PV systems, 
making it necessary to develop scalable models that can handle thousands of 
systems simultaneously, while staying robust to real-world data challenges, 
such as the ability to handle missing data.

In this paper we propose a method to estimate the joint probability distribution 
of the power outputs of a fleet of PV 
systems, modeling all relevant correlations in the data---across individual PV systems and across 
time.  As a copula method, we first develop a novel set of nonlinear marginal transforms that map the 
power from each system to a scalar Gaussian, and then develop a set of linear transformations that 
model the marginally transformed data as a large joint Gaussian distribution. 
We interact with that model to carry 
out various applications including synthetic data generation, data imputation, anomaly detection, 
and forecasting.

\section{Prior work}\label{s-prior-work}
\paragraph{Fleet models} Modeling PV systems for operations and maintenance (O\&M) 
purposes based on measured data 
has a long history~\cite{King1997,King1997a,King2004,Marion2005,DeSoto2006,Holmgren2015}. 
These techniques focus on predicting either the maximum power point or the full 
current-voltage relationship of a PV system under a given set of environmental 
conditions.  O\&M tasks are then carried out using the model. Fault detection is performed by 
comparing actual system power generation to the 
predicted power from the model. Forecasting future PV system power output is done by running 
the models on predicted weather trends, such as those generated by NOAA~\cite{Miller1981}. 
`Fleet modeling' is the practice of constructing an independent, bespoke model for each system in a 
PV fleet and is quite labor intensive. More recently, researchers have attempted to reduce 
the effort of fleet modeling using systematic 
approaches such as learning algorithms and machine inference~\cite{Quintarelli_2019}. 
These approaches tend to be more task dependent, \eg, focusing on the task of anomaly 
detection~\cite{Zhao2019,Aziz2020,Veerasamy2021} or 
forecasting~\cite{Li2020,Huang2020,Ding2021,Plessis2021,Zhao2021,Li2021,Najibi2021}.
While these methods reduce the human effort to create a 
PV system model, they do not provide 
\emph{joint} models of system behavior in a fleet. Several recent papers have explored models to 
predict aggregate quantities of fleets, such as temporal variability and maximum feed-in 
power~\cite{Hoff2010,Hoff2012,Wirth2015,Marcos2016}. 

\paragraph{Copula models and Gaussianization} Our proposed method draws on previous work 
on both copula models and Gaussianization methods.

Copulas are tools for modeling dependence of several random variables, first 
proposed by Abe Sklar in 1959~\cite{Sklar1959} and recently translated into 
English in~\cite{Vliet2023}. 
The basic idea is to apply a nonlinear invertible mapping to each component of a 
random variable so it has some standard distribution such as uniform or Gaussian, 
and then model the dependence of 
these transformed variables~\cite{Schmidt2007,Charpentier2007,Patton2009}. 
Copulas have been applied in 
many domains~\cite{Stander2013,Xu2019,Zhao2020,Schafer1999,Udell2020}, including 
PV data analysis~\cite{Munkhammar2017}. 
Copula models may be based on theoretical constructions (\eg, 
the multivariate Gaussian copula) or may be learned directly from 
data~\cite{Charpentier2007}. 
The \emph{quantile transform} is a typical choice for data-driven models 
of the marginal distributions, and various options exist~\cite{Hardle1994,Pagan1999}. 
Our method includes an autoregressive component in the copula model, an approach that has 
been explored by other authors~\cite{Brechmann2015}.

An alternative approach to modeling multivariate joint distributions is 
\emph{Gaussianization}, also called `normalizing flows'.  
These methods seek an
invertible mapping under which the transformed variable has a
standard (jointly) Gaussian distribution~\cite{Chen2000,Huang2020b,Liao2021}.
The transformations are typically built in steps, so the 
transformation is a composition of multiple transformations, with 
the distribution of mapped variables getting closer to Gaussian as 
more layers or steps are added.
We can think of a Gaussian copula as a simple first step in such a 
normalizing flow.  While a normalization method can in principle model any probability
distribution, Gaussian copula models cannot.
On the other hand a Gaussian copula model makes several operations such as conditioning
on some known values very easy, involving just basic linear algebra.
(These computions can be done for more complex normalizing flows, but they are
much more involved, \eg, requiring Monte Carlo or other sampling type methods.)

\paragraph{Modeling via convex optimization} Our method relies on convex optimization in every 
step. This guarantees efficient algorithms for finding global solutions~\cite{cvx_book}, and mature 
tools exist to easily specify convex optimization problems in 
code~\cite{diamond2016cvxpy,agrawal2018cvxpy}. Many traditional statistical models rely on 
convex optimization for fitting such as regression~\cite[Chap.~12]{Boyd2018}, auto-regressive (AR) 
models~\cite[Ch.~13]{Boyd2018}, and fitting Gaussian 
distributions to data~\cite[\S3.5]{cvx_book}, among many others. Our method is inspired by recent 
work on the trade-off of fit versus roughness in stratified Gaussian 
models~\cite{Tuck2020,Tuck2021}, 
as well as work on convex optimization based signal decomposition~\cite{Meyers2023}.

\section{Data}
\label{s-data}
\begin{figure}
    \centerline{\includegraphics[width=0.6\columnwidth,keepaspectratio,clip=true]{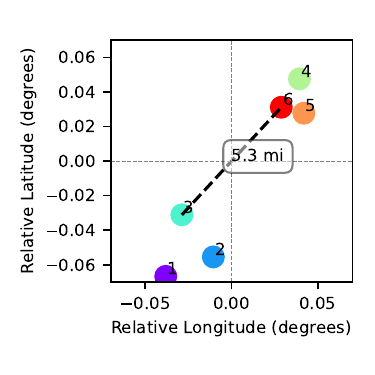}}
    \caption{Relative location of the six PV systems.}
    \label{f-locations}
\end{figure}

We will illustrate our method on PV fleet data provided by 
SunPower Corporation under a nondisclosure agreement.  We select six PV systems 
located in Southern California, with three grouped in Santa Ana, CA and 
three grouped in the hills to the east 
in Tustin, CA. The relative locations are shown in figure~\ref{f-locations}.
This choice of system locations was intentional, as we wanted to verify that
our model captures similarity of power profiles for nearby systems.

\begin{figure*}
    \centerline{\includegraphics[width=\textwidth,clip=true]{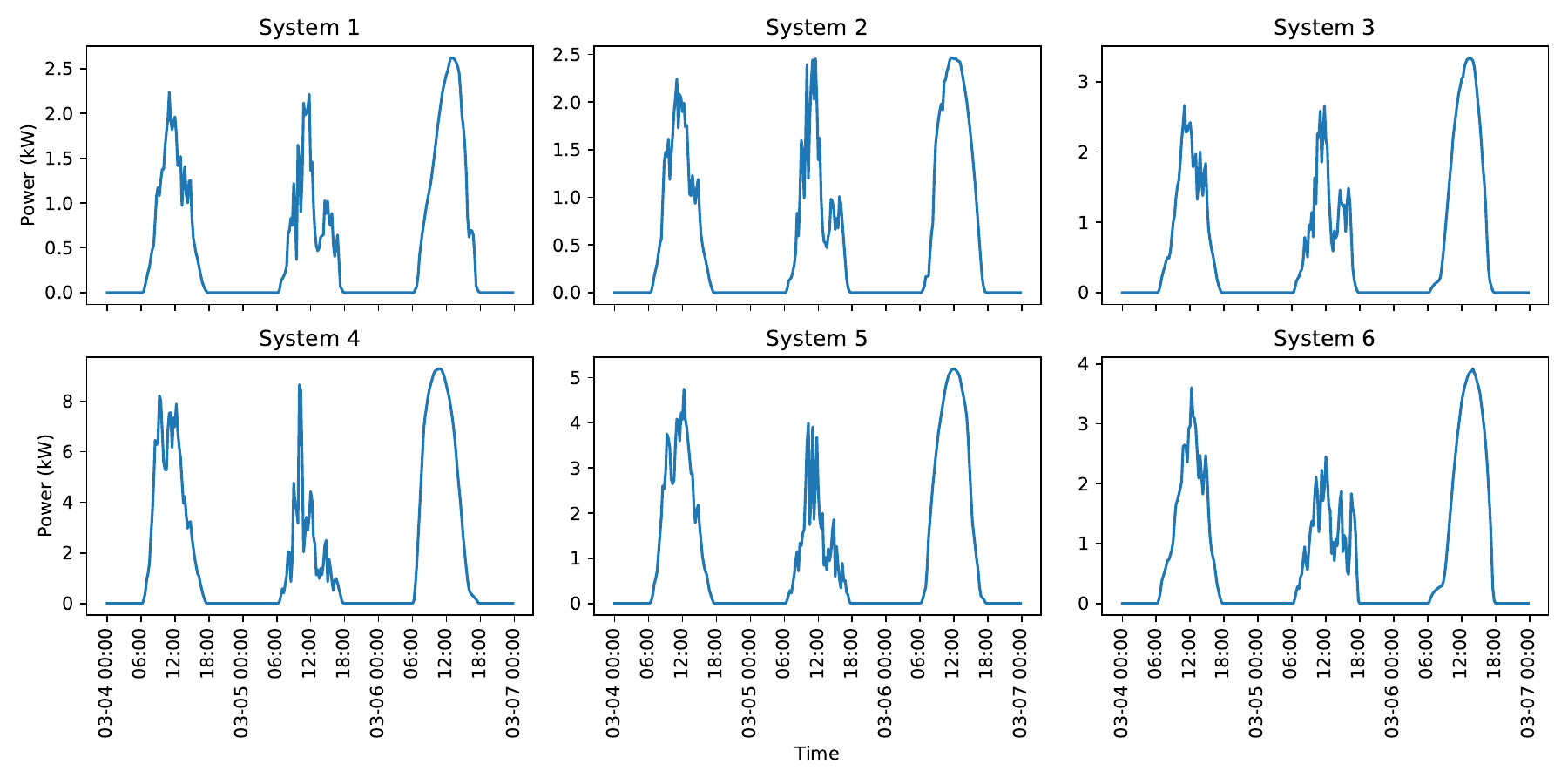}}
    \captionsetup{width=\textwidth}
    \caption{15 minute power output data for six PV systems
from 3/4/2017 to 3/6/2017.}
    \label{f-data}
\end{figure*}

The data consist of 15-minute (average) power values (in kW)
for each of the six systems, recorded  from 3/1/2017 to 3/31/2017.
Figure~\ref{f-data} depicts the power output of the six systems 
over the three day period between 3/4/2017 and 3/6/2017.  
At night the power output is zero;
during daylight hours, we see different types of power profiles.
On 3/6/2017 we see clear-sky behavior, characterized 
by a smooth increase until noon followed by a gradual decrease until evening. 
On 3/5/2017, however,
we see the power generation curves with multiple dips and peaks 
throughout the day, which can be attributed to weather factors such as 
passing clouds.
The maximum power output of the systems 
varies, with system 4 peaking at around 9 kW, and the other systems
peaking at around 2 kW.

We denote the data as $y_t \in \reals^d$, with $d=6$, and
the time index running from $t=1$ to $t=T=2976$.  
This particular data set does not have any missing data.
However, our method gracefully handles missing values, and indeed,
relies on this ability to choose hyper-parameters by cross-validation.

We also use data for the following 2 weeks, from 4/1/2017 to 4/14/2017
as our test set for validating our models and applications.
This data was not used to fit our model.
The test data set has index running from $t=1$ to $T^\text{test}=1344$.

\section{Method}\label{s-method}
\begin{figure}
    \centering
    \begin{tikzpicture}[node distance=3.2cm,
        every node/.style={rectangle, draw, minimum width=1cm, minimum height=0.5cm}]
        \node (box1) {Marginal transform};
        \node (box2) [right of=box1] {AR model};
        \node (box3) [right of=box2] {Residual fit};
        
        \draw[<->,ultra thick] (box1) -- (box2);
        \draw[<->,ultra thick] (box2) -- (box3);
    \end{tikzpicture}
    \caption{Three invertible tranformations used in the proposed method.}
    \label{f-models}
\end{figure}
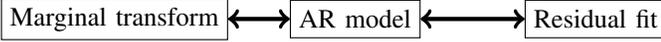

We propose a method for fitting the given data $y_1, \ldots, y_T$ 
to a smooth 24-hour-periodic stochastic process.
We apply a sequence of three invertible 
transformations so that the transformed data is approximately a standard Gaussian.
These transformations, which respect periodicity and are constructed to vary 
smoothly across time, are applied in the three steps shown in 
figure~\ref{f-models}. 
First we use a smooth periodic nonlinear transform to make the
data approximately marginally Gaussian. This allows us to model the changing distribution over a 
day, in addition to the differing maximum values seen across systems.
In the second step we use an autoregressive (AR) model to account for dependencies
across time.  This results in a residual that is approximately uncorrelated across time.
Finally, we fit a smooth periodic Gaussian distribution to the residual of 
the AR model; from this we can whiten the residual so that it is 
approximately a standard Gaussian. 
In the language of copula modeling, our first step is our marginal transformation
and the final two steps constitute our copula (\ie, `linking') function.

\subsection{Fitting a smooth periodic model} \label{s-fitting-sp}
Here we describe the general technique, used in steps~1 and~3 of our method,
for fitting a smooth $P$-periodic parameter, given by 
$\theta_1, \ldots, \theta_T\in
\Theta \subseteq \reals^m$ to some data, where $\Theta$ is a convex 
set of allowed parameter values. 
We will use a Fourier series with $K$ harmonics to represent $\theta$,
\BEQ \label{e-fourier-series}
\theta_t= \sum_{k=0}^{K} \left( \cos\left(\frac{2\pi k t}{P}\right) \alpha_k
+ \sin\left(\frac{2\pi k t}{P}\right) \beta_k \right),
\EEQ
for $t=1, \ldots, P$,
where $\alpha_{k},\beta_{k} \in \reals^m$ are the (vector) coefficients that define
$\theta$.

We take as a measure of smoothness the Dirichlet energy,
\BEAS
\mathcal D &=& \frac{(2\pi)^2}{P} \sum_{k=1}^K k^2\left(\|\alpha_k\|_2^2 +
\| \beta_k\|_2^2 \right).
\EEAS
The loss function has the form
\[
\mathcal L = \sum_{t=1}^T \ell_t (\theta_t),
\]
where $\ell_t: \Theta \to \reals$ is a convex loss function
that depends on some data at time $t$. (If some data are missing, 
then the sum is only over $t$ for which the data are available.)

Our generic fitting method takes $\theta$ as a solution of the convex 
 optimization problem
\BEQ \label{e-sp-fit}
\begin{array}{ll} 
\mbox{minimize} & \mathcal L + \lambda \mathcal D,\\
\mbox{subject to} & \theta_t \in \Theta, \quad t=1, \ldots, P,
\end{array}
\EEQ
where $\lambda>0$ is the smoothing regularizer hyper-parameter.
The variables are the $m$-vectors 
$\alpha_0, \ldots, \alpha_K$ and $\beta_1, \ldots, \beta_K$.
This is a convex optimization problem, and readily solved.

This generic fitting method contains the hyper-parameter $\lambda$
(and possibly others),
but good values of these can be found automatically 
using cross-validation~\cite[\S7.10]{Hastie2009},\cite[\S13.2]{Boyd2018}, 
so the method is essentially hyper-parameter free and automatic.

Two steps of our method solve a different instance 
of the problem \eqref{e-sp-fit} with $\lambda$ chosen using 
cross-validation. 
This method of fitting a smooth periodic parameter is a special
case of a Laplacian regularized stratified model~\cite{Tuck2020,Tuck2021}
with the underlying graph a cycle representing the periodicity. 
It can also be thought of as a signal 
decomposition problem~\cite{Meyers2023}.

\subsection{Marginal transforms}\label{s-marginal-transforms}
In this first step, we seek continuous increasing functions 
$\varphi_{t,i}: \reals \to \reals$. We allow these functions to change with $t$ but enforce that they 
are periodic and smooth. Our goal is for the transformed values,
\[
x_{t,i} = \varphi_{t,i} (y_{t,i}) , \quad~\mbox{for}~t=1,\dots,T,
\]
to have an approximately Gaussian (marginal) distribution for each $i=1, \ldots, d$.
Typical quantile transforms are static; but
our method defines a transformation that changes smoothly in time and is periodic.

We carry out this step for each component of the 
original data, so the method described in this section
is carried out separately for each $i=1, \ldots, d$.
To lighten the notation in this section, we drop the component index $i$,
and consider the original data $y_t$ to be scalar.

Our first step is to estimate a set of quantiles 
$0\leq \eta_1< \cdots < \eta_r \leq 1$ of the data.
By default we take these to be 2nd percentile, the 98th percentile, and 
the 10 deciles, so $r=11$ and
\[
\eta = (0.02, 0.10, 0.20,\ldots, 0.80, 0.90, 0.98),
\]
but our method is general and any other choice of quantiles could be used.
We denote the estimated $\eta_i$ quantile of $y_t$ as 
$q_{t,i}$, $t=1, \ldots, T$,
$i=1,\ldots, r$.  We assume these are $P$-periodic and smooth.

To estimate these quantiles from the data we use standard quantile 
regression~\cite{Koenker2005,Koenker2017},
which relies on the so-called
pinball loss, defined as
\[
\ell^\mathrm{pin}(u; \eta) =  \max\{ (1-\tau)u, \tau u\} = (\tau-1/2) |u| + (1/2)u,
\]
for quantile $\tau \in [0,1]$.
We take our loss function to be
\BEQ\label{e-quantiles-loss}
\ell (q_t) =
\ell^\mathrm{pin}(q_{t,1}-y_t; \eta_1) + \cdots +
\ell^\mathrm{pin}(q_{t,r}-y_t; \eta_r),
\EEQ
the sum of the pinball losses associated with each of our $r$ quantiles.
To estimate the quantiles we solve the generic problem \eqref{e-sp-fit},
with loss function \eqref{e-quantiles-loss}, and constraint set
\[
\Theta = \{ q \mid q_{1} \leq \cdots \leq q_{r} \},
\]
which enforces that the quantiles are consistent.
This simultaneously estimates the $r$ quantiles of $y_t$ for each $t$,
with the quantiles being smooth and periodic, and always satisfiying
the consistency constraint.
The hyper-parameter $\lambda$, which controls how smooth the quantile estimates are,
can be chosen automatically via cross-validation.

Given these periodic smooth quantile estimates, 
we construct nonlinear mappings $\varphi_t: \reals\to\reals$
as continuous piecewise linear functions, with knot-points
given by the quantiles, and values at those points given by the 
associated value of a standard scalar Gaussian for the same quantiles, \ie,
\BEQ\label{e-phi-def}
\varphi_t(q_{t,j}) = \Phi^{-1}(\eta_j), \quad j=1, \ldots, r,
\EEQ
where $\Phi$ is the cumulative distribution function (CDF) of a standard Gaussian.
This gives the marginally transformed data 
$x_{t,i}=\varphi_{t}(y_{t,i})$.

\subsection{Autoregressive model}\label{s-AR-fit}
The time series $x_1, \ldots, x_T$ has entries with approximately standard Gaussian
marginal distribution, but there are dependencies between the components,
as well as across time.
Our next step is to handle the dependency across time.
We fit a vector autoregressive (AR) model to the 
marginally Gaussianized data $x_t$.
The model is
\BEQ \label{e-var}
x_{t} = A_{1} x_{t-1} + \cdots + A_{M}x_{t-M} +  v_{t},
\EEQ
where $v_{t}$ is a process noise or residual, $M$ is the memory of the AR model,
and $A_{1}, \ldots, A_{M} \in \reals^{d \times d}$ are the coefficients.
We could fit these AR coefficients as smooth and periodic, but we have found
that a constant AR model
does just as well as a more complex time-varying one. 

We fit these coefficients by minimizing the mean-squared error,
the average of
\[
\ell_t = 
\left\| x_{t} - A_{1} x_{t-1} + \cdots + A_{M}x_{t-M} \right\|_2^2
\]
over those entries where all $x_t$ are known.
We add ridge regularization to this average loss,
\[
\lambda^\text{ridge} \left( \|A_1\|_F^2+ \cdots + \|A_M\|_F^2 \right),
\]
where $\lambda^\text{ridge}>0$ is a hyper-parameter that scales the regularization,
and $\| \cdot \|_F^2$ denotes the square of the Frobenius norm, \ie,
the sum of squares of entries.
The hyperparameter $\lambda^\text{ridge}$ can be chosen by cross-validation.

We denote the AR residual as
\[
v_t = 
x_{t} - (A_{1} x_{t-1} + \cdots + A_{M}x_{t-M}), 
\]
defined when $x_t, \ldots, x_{t-M}$ are all known.
Note that in the special case $M=0$, which corresponds to the 
model that $x_t$ are approximately uncorrelated for different $t$,
the residual reduces to $x_t$.

\subsection{Smooth periodic residual fit}
Our last step is fit a smoothly varying periodic Gaussian distribution to the residual $v_t$, 
$v_t \sim \mathcal N(\mu_t, \Sigma_t)$, where we assume that $v_s$ and $v_t$ 
are independent for $s \neq t$.
We model $v_t\in \reals^d$ as smooth periodic Gaussian,
\BEQ \label{e-gaussian-xt}
v_t \sim \mathcal{N} (\mu_t, \Sigma_t), \quad t=1,\ldots,T
\EEQ
where $\Sigma_t$ and $\mu_t$ are smooth and periodic. 
We expect $\mu_t$ to be small.

Our loss $\ell_t$ will be the negative log-likelihood of the Gaussian 
model \eqref{e-gaussian-xt}, which is
\[
\ell_t = \frac{d}{2} \log (2\pi) - \sum_{j=1}^{d} \log \left(\diag \left( L_{t} \right)_{j} \right)
+ \frac{1}{2} \| L_t^T v_t - \nu_t \|_{2}^{2},
\]
with variables $L_t \in \reals^{d \times d}$ and $\nu_t \in \reals^d$. 
Here, $L_t$ is the Cholesky factor of $\Sigma^{-1}_t$ 
and $\nu_t = L_t^{-T} \mu_t$.
This change of variables makes the loss a convex function. 
We can recover $\Sigma_t$ and $\mu_t$ as
\[
\Sigma_t = (L_tL_t^T)^{-1}, \qquad
\mu_t =  L_t^{-1} \nu_t.
\]
once we solve the optimization problem to find $L_t$ and $\nu_t$.
Here too the smoothness hyper-parameter $\lambda$ is found by
cross-validation.

Our final whitened signal is given by
\[
z_t = \Sigma_t^{-1/2} (x_t-\mu_t) = L_t^T x_t-\nu_t,
\]
defined when $x_t$ is.
According to our model, these are independent identically distributed (IID)
with $z_t \sim \mathcal N(0,I)$.

\subsection{The whole model}
From our first two steps,
we see that our model of $y_t$ is a stationary periodic 
Gaussian process $x_t$, mapped entrywise through a smoothly periodic 
transformation.  Using all three steps, we interpret it as IID 
Gaussians $z_t$, passed through 
an AR filter to obtain $y_t$, and then mapped entrywise.

Such a model allows us to carry out several operations.
We can generate samples from the model. We can evaluate
the density at a sequence $y_t$.
We can condition on a set of known values of some of the 
components, as well as computing conditional marginal quantiles
for each unknown entry.
These allow us to carry out imputation, \ie, guessing 
missing values, by evaluating the conditional median of a missing
entry given the known ones.
(We also can get error bars, \eg, the 10th and 90th conditional
marginal quantiles.)
We can also do anomaly detection, where we 
detect known entries that do not fit the model.
To do this we compute the 
conditional quantile of each known entry, given the other known entries but not
that particular value;
conditional quantile values that are either near zero or one are then 
flagged as suspicious.

These operations (and others) 
can be carried out for many types of statistical models, 
for example using Monte Carlo sampling.
But due to the specific structure of our model, we can carry them out
using simple linear algebra, which makes the operations fast
and reliable.
Details of how we implement these operations will be given 
in a forthcoming paper.

\section{Results}
\label{s-results}
\begin{figure}
    \centerline{\includegraphics[width=\columnwidth,clip=true]{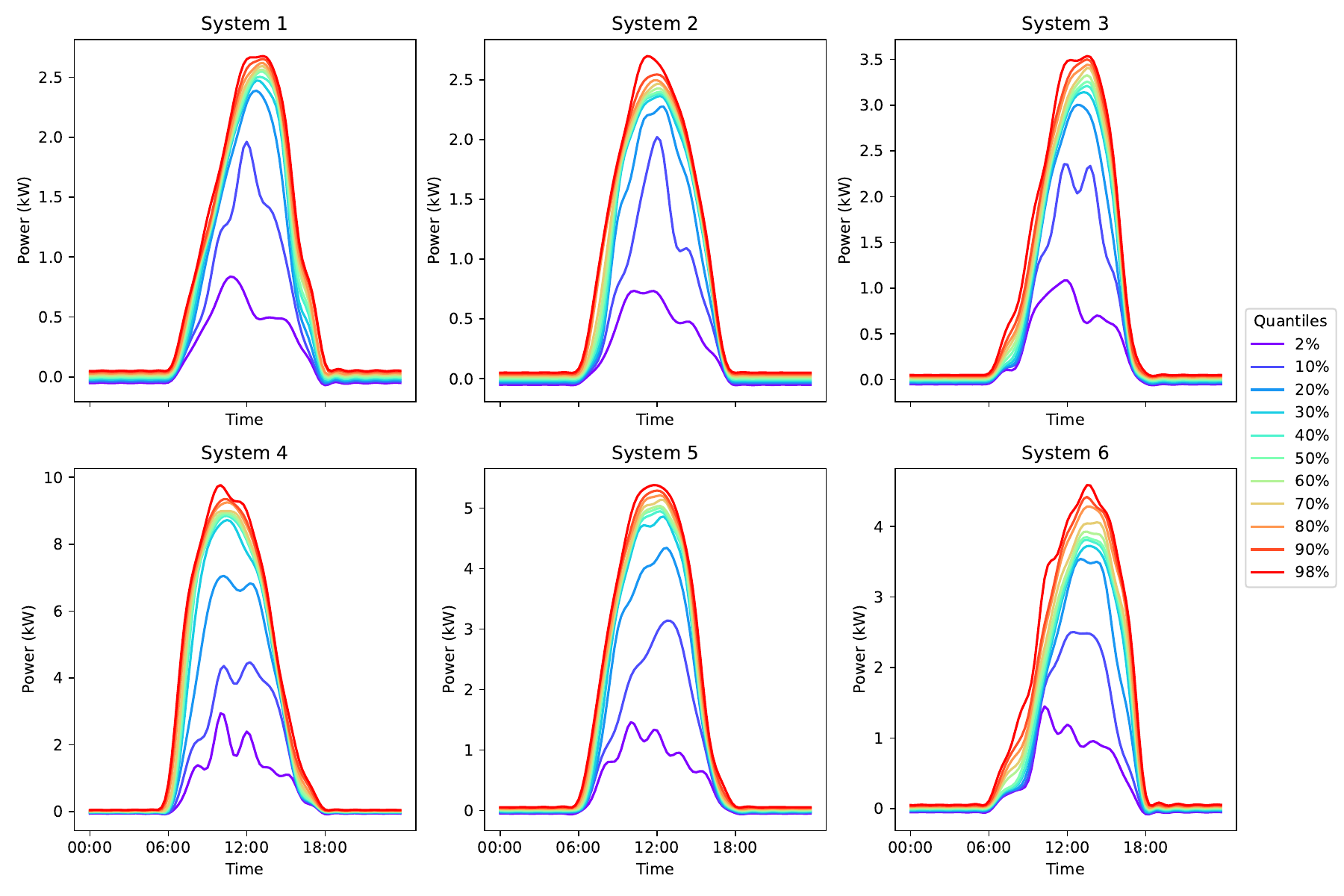}}
    \caption{Smooth periodic quantiles.}
    \label{f-quantiles}
\end{figure}

Here we show the results of our modeling method on the PV data 
described above, using default parameters.
Estimated quantiles for each system are shown in figure~\ref{f-quantiles}. 
Note that the 98th percentile serves as an effective statistical clear-sky model.
The estimated quantiles collapse to zero at night as expected. 
The spread between the upper quantiles is narrower than that of the lower quantiles,
especially around noon. 
Quantiles for each system exhibit distinct characteristics, with
the systems physically near each other showing similar shaped quantiles.

\begin{figure}
    \centerline{\includegraphics[width=\columnwidth,clip=true]{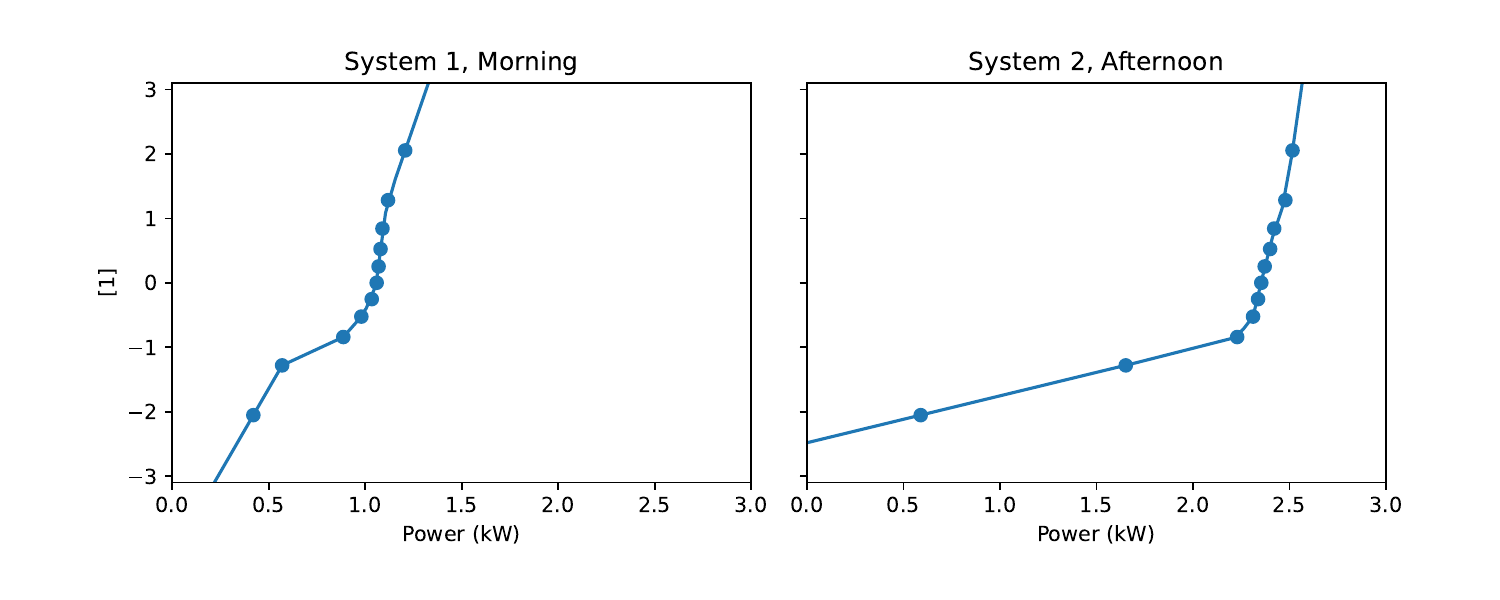}}
    \caption{Time aware copula transform.}
    \label{f-pwl}
\end{figure}

A few samples of the associated piecewise linear copula transforms 
are shown in figure~\ref{f-pwl}. We see that the same power value 
is mapped to different points based on the time of day and 
the system. This shows how our time-aware 
copula transform adapts to both the time of day and the 
unique characteristics of different systems as 
opposed to a standard fixed copula transform, which does not.

We use AR memory $M=3$, with coefficients.
We observe several interesting phenomena in these coefficient matrices.
First, the entries of $A_1$ are generally bigger than those of $A_2$ and $A_3$,
showing that the previous period plays a larger role in predicting the current values
than the previous two values.  We also see that the diagonal entries are generally 
larger than the off-diagonal ones, meaning that the previous values for 
each system play a larger role in predicting the current value than the previous values
of the other systems.
However, the many non-zero off-diagonal elements in the coefficient matrices show
that the predictions for each system do depend on the previous values
of the other systems.

\begin{figure}
    \centerline{\includegraphics[width=0.9\columnwidth,clip=true]{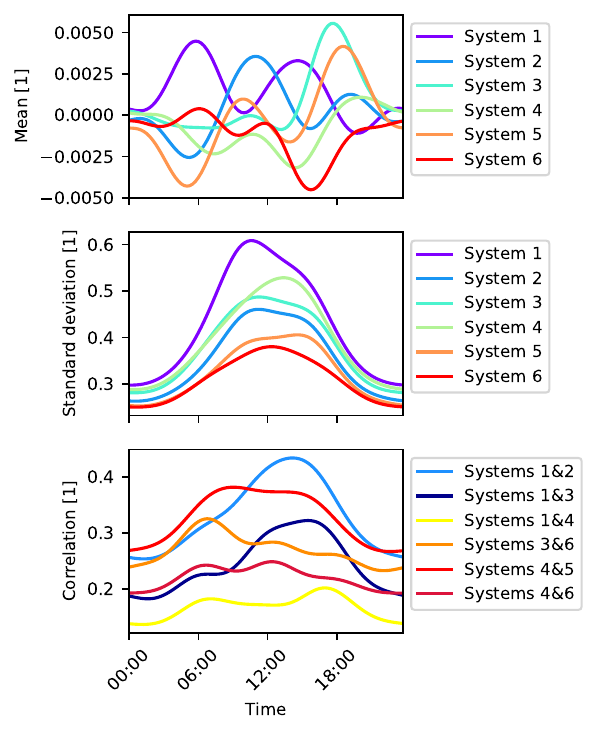}}
    \caption{Means, standard deviations, and selected correlations of 
AR residuals.}
    \label{f-residuals}
\end{figure}

Finally we fit a smoothly varying periodic Gaussian distribution to 
the residuals of the AR model, shown in figure~\ref{f-residuals}.
The top plot shows the means, which are indeed small, as expected.
The middle plot shows the standard deviations of the residuals.  These are smaller 
than one, which is approximately the standard deviation of the entries of $x_t$,
which means we are able to predict the current values using previous values
better than simply guessing $x_t=0$, \ie, treating them as uncorrelated across
time.
We can see that the residual standard deviations
vary considerably across systems and time of day.  Roughly speaking,
the residual of system~1 has almost twice the standard variation of the
residual of system~6.  We also see that the
residual standard deviation is smaller at noon than in the morning and afternoon.
The bottom plot shows the correlations of selected pairs of residuals.
These correlations are generally around 30\%, but we can see that systems 
that are physically near each other are more highly correlated.  We can also
see variation of the correlation over the day.

\section{Applications}
\label{s-applications}

\subsection{Generating simulated data}
\begin{figure}
    \centerline{\includegraphics[width=0.7\columnwidth,keepaspectratio,clip=true]{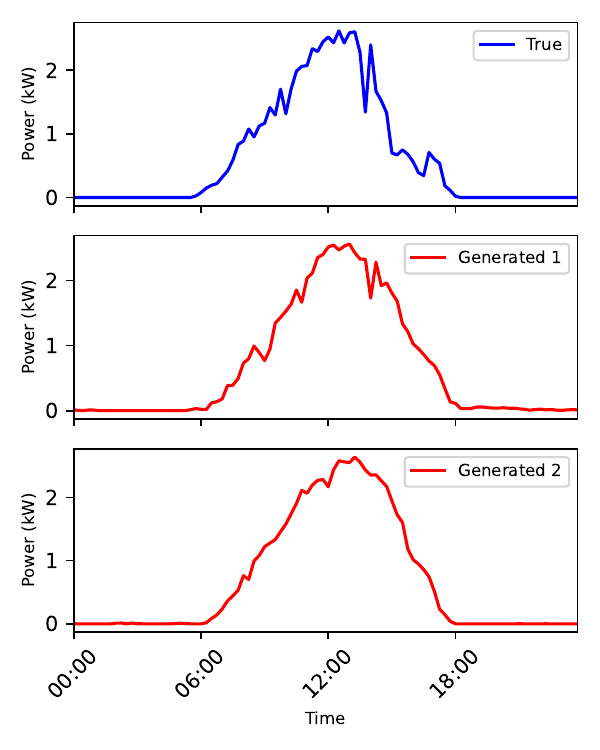}}
    \caption{\emph{Top.} System~1 on 4/7/2017. \emph{Middle and bottom.}
Two simulations from our model.}
    \label{f-simulations}
\end{figure}

We generate simulated data from our model by simulating data from the periodic 
Gaussian stochastic process, and then applying the inverse nonlinearities $\phi_{t,i}^{-1}$ to
the entries of these samples.
Figure~\ref{f-simulations} shows two simulations of fake data for system~1, 
with the actual data for the specific day 4/7/2017 shown at top for reference.  
They appear quite similar.

\subsection{Conditional marginal quantiles}
\begin{figure}
    \centerline{\includegraphics[width=0.7\columnwidth,keepaspectratio,clip=true]{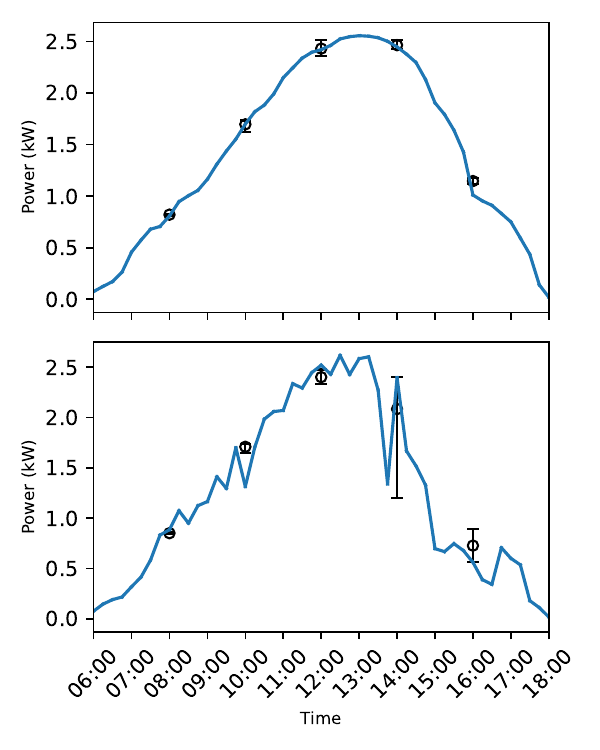}}
    \caption{10th, 50th, and 90th marginal conditional quantiles at five
times, conditioned on values of all systems at other times.  
\emph{Top.} Clear day.
\emph{Bottom.} Cloudy day.}
    \label{f-marginal_quantiles}
\end{figure}

We can compute the marginal quantiles of any entry, conditioned on all other 
known entries, in time and across systems.  When the entry is unknown, this gives us a sophisticated 
method for imputing or guessing what the missing value might have been.  
We can use the conditional marginal 
median (50th percentile) as the imputed value, with the 10th and 90th percentiles defining 
an uncertainty interval.
Figure~\ref{f-marginal_quantiles} shows the marginal conditional quantiles 
for system~1 at 5 times on two 
days, one clear and one partially cloudy.
We observe that the model correctly adapts the uncertainty bounds to the weather, with tighter 
bounds on clear days. Additionally, we note that the uncertainty bounds are asymmetric, with
decreases in output (say, due to clouds) more likely than increases.
The predictions themselves, shown as the circle representing
the conditional median, are good.

\subsection{Anomaly detection}
We can use marginal conditional quantiles to identify anomalous entries in our data.
To estimate whether a given known entry is an anomaly we pretend that it is unknown,
compute its conditional marginal CDF given all other known values,
and evaluate it at the known value.
We can flag an entry as anomalous if this quantile value is less 
than $\epsilon$ or more
than $1-\epsilon$, where $\epsilon$ is a threshold value such as $10^{-2}$.
With this threshold, we would expect a false positive rate around $2\epsilon$.

To illustrate this, we consider system~2 on 4/1/2017.
We introduce synthetic anomalies by perturbing the power values at
8:30, 10:00, 11:30, 13:00, 14:15 and 15:30, by randomly 
increasing or decreasing the true values by 15\%.
Table~\ref{t-anomaly} shows true values,
perturbed values and conditional marginal quantiles of perturbed values,
clipped to the range $[0.0001,0.9999]$.

\begin{figure}
    \centerline{\includegraphics[width=0.7\columnwidth,keepaspectratio,clip=true]{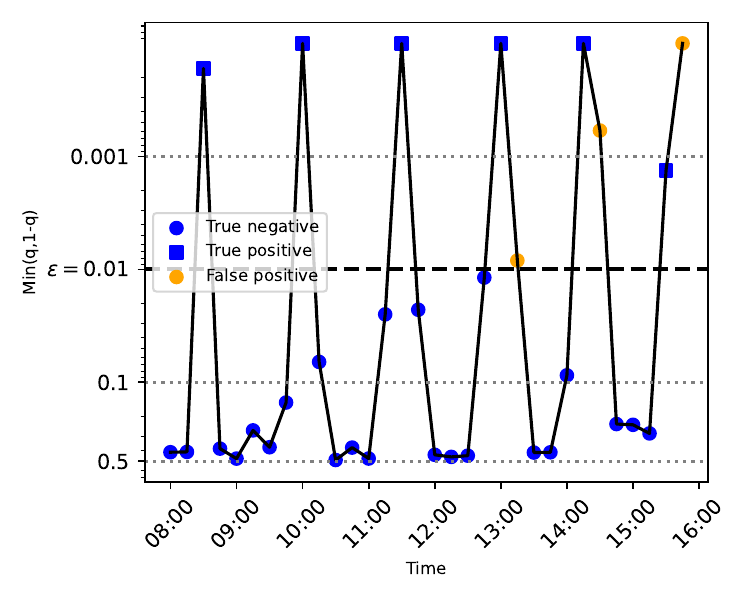}}
    \caption{Marginal conditional quantiles for system~2 on 4/1/2017, 
with six artificial anomalies added.}
    \label{f-anomaly_detection}
\end{figure}

With threshold $\epsilon = 0.01$, we detect all of the artificial
anomalies. 
We also have three false positives, \ie, times 
when a true value is flagged as an anomaly.
The conditional marginal quantiles for this day 
are shown in figure~\ref{f-anomaly_detection}.
The vertical axis shows $\min\{q,1-q\}$, with the threshold $\epsilon=0.01$
shown as the darker horizontal line.
True negatives are shown as blue circles, and true positives are shown
as blue squares.  False positives are shown as orange circles.
The three false positives are all cases where the true power was 
low compared to our predicted median.  This is not suprising; 
clouds can easily reduce power output unexpectedly by 15\% or more.
\begin{table}
    \caption{Anomaly detection example.}
    \begin{center}
    \begin{tabular}{|c|c|c|c|}
    \hline
    Time & True value & Perturbed value & Conditional quantile \\
    \hline
    08:30 & 0.9919 & 0.8431 & 0.0001 \\
    \hline
    10:00 & 1.7280 & 1.9872 & 0.9999 \\
    \hline
    11:30 & 2.3633 & 2.7178 & 0.9999 \\
    \hline
    13:00 & 2.5890 & 2.9773 & 0.9999 \\
    \hline
    14:15 & 2.4294 & 2.0650 & 0.0001 \\
    \hline
    15:30 & 1.6576 & 1.4090 & 0.0013 \\
    \hline
    \multicolumn{4}{l}{}
    \end{tabular}
    \label{t-anomaly}
    \end{center}
\end{table}

\subsection{Forecast}
\begin{figure}
    \centerline{\includegraphics[width=0.7\columnwidth,keepaspectratio,clip=true]{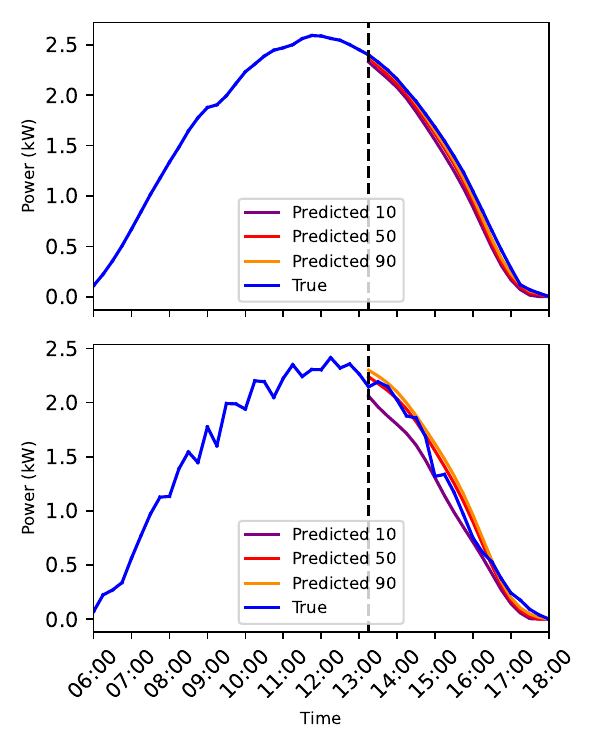}}
    \caption{Forecast for system~2 on the clear day 4/9/2017, and the cloudy day 4/6/2017. We forecast
from 13:15 on, given values for all systems up through 18:00.}
    \label{f-forecasts}
\end{figure}

Here we forecast the values of system~2 from 13:15 on, using data from all
systems up through 13:00.  We show the forecast, which is the median or 50th 
conditional marginal quantile, along with the conditional marginal 10th and 90th 
quantiles, which give us confidence bands for the forecast values.
This is illustrated in figure~\ref{f-forecasts} on
the clear day 4/9/2017 and the cloudy day 4/6/2017.
Our forecast on the clear day is very good, with tight uncertainty bands.
Our forecast on the cloudy day is reasonable, but with much wider uncertainty bands.

\begin{figure}
    \centerline{\includegraphics[width=0.7\columnwidth,keepaspectratio,clip=true]{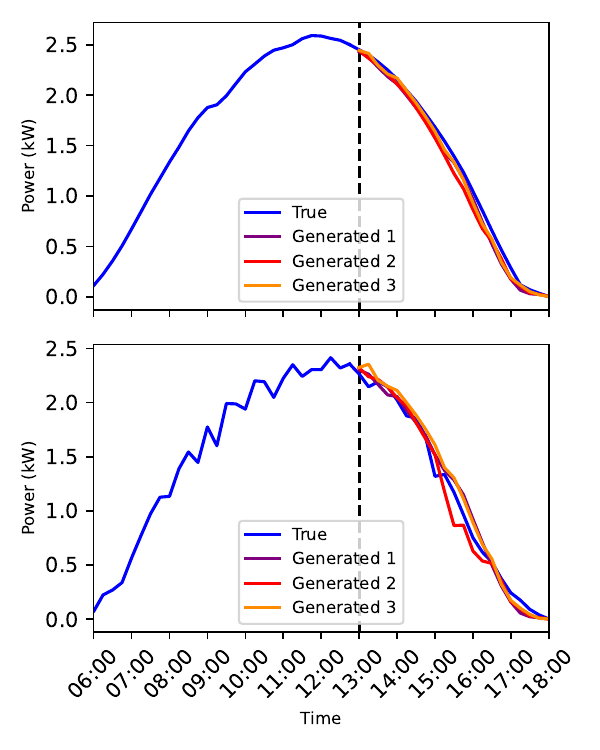}}
    \caption{Forecast for system~2 on the clear day 4/9/2017, and the cloudy day 4/6/2017. We forecast
from 13:15 on, given values for all systems up through 18:00.}
    \label{f-joint_forecasts}
\end{figure}

In addition to forecasting marginal quantiles of a single system, we can generate
joint forecasts using all systems.  We illustrate this in figure~\ref{f-joint_forecasts},
where we forecast the values of system~2 from 13:15 on, using data from all 
systems up through 13:00.  We show three different forecasts, sampled from
the full joint conditional distribution.  We see that the forecasts agree 
with the marginal forecasts, in the sense that generated instances are within the 
10--90\% confidence bands of the marginal forecasts. We also see that the forecasts
are reasonable, since we observe that for cloudy days, the forecasts are more volatile
than for clear days with the latter having a more stable, smooth behavior.

\section{Conclusions}
We presented a novel approach to
modeling and analyzing fleets of PV systems based on 
a smooth periodic Gaussian copula transform, and illustrated some of its
applications. 
While we have demonstrated the method on a small example, it can scale
gracefully to much larger problems; details will be given in a forthcoming paper.

\section*{Acknowledgment}
This material is based on work supported by the U.S. Department of Energy's Office of
Energy Efficiency and Renewable Energy (EERE) under the Solar Energy Technologies Office
Award Number 38529. Stephen Boyd's work was funded in part by the AI Chip Center for
Emerging Smart Systems (ACCESS).
\small
\bibliographystyle{IEEEtran}
\bibliography{references}
\vspace{12pt}
\color{red}

\end{document}